\documentclass[aps,pre,preprint,floatfix]{revtex4}

\usepackage{latexsym}
\usepackage[usenames]{color}
\usepackage{amsthm}
\usepackage{hyperref}

\usepackage{amsmath}    % need for subequations
\usepackage{amsfonts}  %note how statements can be commented out
\usepackage{amssymb}
\usepackage{graphicx}
\usepackage{slashed}
\usepackage{fouridx}

\theoremstyle{plain} 
\theoremstyle{plain} 
\theoremstyle{plain}  
\theoremstyle{plain} 
\theoremstyle{plain}  
\theoremstyle{remark} 
\theoremstyle{plain} 
\theoremstyle{remark}

\newcommand\CROSS[1]{%
  \hbox{%
    \vbox{
      \hrule
      \kern2.5pt
      \hbox{$#1$\,\strut}
    }%
  \vrule
  }\mskip\thickmuskip
}

\begin{document}
\begin{center}
\Large{\textbf{Fractal surfaces from simple arithmetic operations}}\\ 
~\\

\large{Vladimir Garc\'{\i}a-Morales}\\

\normalsize{}
~\\
%Institute for Advanced Study -  Technische Universit\"{a}t M\"{u}nchen,\\ Lichtenbergstr. 2a, D-85748 Garching, Germany \\

Departament de Termodin\`amica, Universitat de Val\`encia, \\ E-46100 Burjassot, Spain
\\ garmovla@uv.es
\end{center}
\small{}
\noindent Fractal surfaces ('patchwork quilts') are shown to arise under most general circumstances involving simple bitwise operations between real numbers. A theory is presented for all deterministic bitwise operations on a finite alphabet. It is shown that these models give rise to a roughness exponent $H$ that shapes the resulting spatial patterns, larger values of the exponent leading to coarser surfaces.
~\\

\noindent keywords: fractals; self-affinity; free energy landscapes; coarse graining
\pagebreak

%\small{ \\
%%vmorales@ph.tum.de}

\section{Introduction}
\label{intro}
%\subsection{Generalized bitwise arithmetic}

Fractal surfaces \cite{Mandelbrot, Feder} are ubiquitously found in biological and physical systems at all scales, ranging from atoms to galaxies \cite{Moriarty, Richardella, Martinez, Baryshev}. Mathematically, fractals arise from iterated function systems \cite{PeitgenBOOK, Barnsley}, strange attractors \cite{Ott}, critical phenomena \cite{SornetteBOOK}, cellular automata \cite{Wolfram, VGM1}, substitution systems \cite{Wolfram, VGM4} and any context where some hierarchical structure is present \cite{Yamamoto, Yamamoto2}. Because of these connections, fractals are also important in some recent approaches to nonequilibrium statistical mechanics of steady states \cite{Tsallis, Tsallis2, Hoover, VGMStat}. 

Fractality is intimately connected to power laws, real-valued dimensions and exponential growth of details as resulting from an increase in the resolution of a geometric object \cite{Indekeu5}. Although the rigorous definition of a fractal requires that the Hausdorff-Besicovitch dimension $D_{F}$ is strictly greater than the Euclidean dimension $D$ of the space in which the fractal object is embedded, there are cases in which surfaces are sufficiently broken at all length scales so as to deserve being named `fractals' \cite{Mandelbrot, Indekeu5, MVBerry}. Most of these surfaces are random and governed by growth, adsorption and deposition processes on interfaces \cite{Indekeu5,Vicsek, Vicsek2, StanleyFractals, Indekeu2, Indekeu3, Indekeu4, Indekeu6}. 

\begin{figure*} 
\includegraphics[width=1.0 \textwidth]{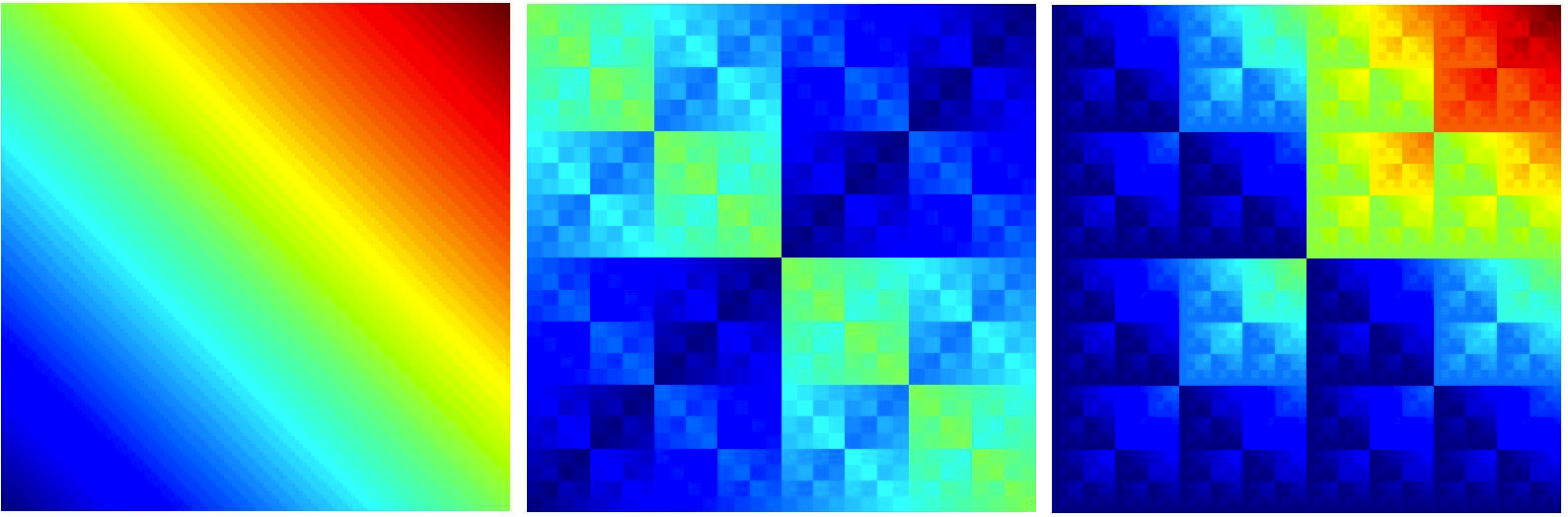}
\caption{\scriptsize{The functions $f(x,y)=x+y$ (left), $g(x,y)=x+_{2}y$ (center) and $h(x,y)=x\hat{+}_{2}y$ (right) for $x \in [0,1]$ and $y \in [0,1]$. We have $f(x,y)=g(x,y)+h(x,y)$ everywhere. Dark blue means a zero value, green a value of 1 and dark red a value of 2.}} \label{dsums}
\end{figure*}

The multidisciplinary field of \emph{disorderly} surface growth has experienced a rapid development \cite{StanleyFractals}. However, fractal surfaces formed by \emph{deterministic} processes and hierarchical rules are also interesting and can be useful, e.g., as models for cityscapes \cite{Indekeu5}. In this article, by regarding `fractals' in the broad sense mentioned in the previous paragraph, we construct a wide variety of such surfaces and show that they possess fractal self-affine properties. If a surface $F(x,y)$ is self-affine it satisfies  \cite{Vicsek, Vicsek2}
\begin{equation}
F(x,y) \sim b^{-H}F(bx,by) \label{selfafi}
\end{equation}	
where $H \in \mathbb{R}$ is the roughness exponent characterizing the self-affine scaling \cite{Mandelbrot} \cite{Sinha}. We prove that the surfaces here constructed obey Eq. (\ref{selfafi}). Although our construction proceeds abstractly, we illustrate it with specific numerical examples, and we believe that the generality of the method is such that it may find many applications in the modeling of complex physical systems. Although the notation may seem unfamiliar, the mathematics behind is elementary, and is based on generalized bitwise arithmetic on real numbers. To introduce this idea, let us consider two real numbers $a$ and $b$ (that we may truncate to a finite number of digits after the decimal point). If we expand $a$ and $b$ in base 10, we see that the ordinary sum of these numbers splits into the ordinary sum of two different parts
\begin{equation}
a+b=(a+_{10}b)+(a\hat{+}_{10}b)
\end{equation}
here $+_{10}$ denotes addition modulo $10$ and $\hat{+}_{10}$ denotes the contribution of the carries to the sum $a+b$. If $0\le a <10$ and $0\le b <10$ then, it is clear that $a+b=a+_{10}b$ if $a+b <10$. If $a+b \ge 10$ then $a+_{10}b=a+b-10$. If $a$ and/or $b$ are larger than 10, the sum $a+b$ is clearly reduced to separately considering the digits of $a$ and $b$ and adding them, taking care of the carries. Addition modulo 10 means that the carries are neglected. For example, if $a=5.6782$ and $b=3.6754$ we have that $a+b=9.3536$. This sum can be seen as the ordinary addition of two different contributions $a+_{10}b=8.2436$ and $a\hat{+}_{10}b=1.1100$. The structure of the operators $+_{10}$ and $\hat{+}_{10}$ is very interesting. All positions within the numbers are independent of each other as regards the bitwise action of $+_{10}$: Each two digits corresponding to the same power of ten are added modulo ten and no carry is transferred from one position to another. Therefore, under the action of such operator, the positions within the number are `uncoupled' and independent. Since each position within a number in a standard positional number system corresponds to a different power of the base, adding modulo 10 two real numbers means \emph{performing the same operation at all scales, the latter being independent of each other}. We claim that, as a result of this, the function $g(x,y)=x+_{10}y$ (where $x$ and $y$ are real numbers) has fractal features. It is a discontinuous and non-differentiable function that exhibits the same details at all scales, displaying self-similarity. In fact this is generally the case for the function $g(x,y)=x+_{p}y$, where $p \in \mathbb{N}$, $p\ge 2$ is any base, with $+_{p}$ denoting \emph{addition modulo $p$} (\cite{VGM2, VGM3}). \emph{We shall prove below that such a function $g(x,y)$ obeys Eq. (\ref{selfafi})}. Furthermore, since $f(x,y)=x+y$ is not a fractal, the function $h(x,y)=x\hat{+}_{p}y=x+y-(x+_{p}y)$ must be a fractal as well so as to compensate the discontinuities of $g(x,y)$. In Fig. (\ref{dsums}) the functions $f(x,y)=x+y$ (left), $g(x,y)=x+_{2}y$ (right) and $h(x,y)=x\hat{+}_{2}y$ are plotted for $x \in [0,1]$ and $y \in [0,1]$. While $f(x,y)=x+y$ is just an inclined plane, the functions $g(x,y)=x+_{2}y$ and $h(x,y)=x\hat{+}_{2}y$ exhibit a more complex behavior displaying `squares within squares' at all scales. The behavior of these functions is such that under ordinary addition, all discontinuous jumps disappear at all scales, yielding the smooth function $f(x,y)$ (which is \emph{everywhere} equal to $g(x,y)+h(x,y)$).

The purpose of this paper is to construct \emph{all possible bitwise operators acting on a finite alphabet of $p$ symbols} (to which $+_{p}$ and $\hat{+}_{p}$ do belong, and which we explicitly construct as well, as an example) and to elucidate their associated fractal features. We thus find a new kind of surfaces with self-affine properties, that we term `patchwork quilts'. We constructively prove the general fact that bitwise operators locally acting on each point $(x,y)$ in the plane $\mathbb{R}^{2}$ leads to these surfaces, providing explicit mathematical expressions for them. The results can be easily extended to hypersurfaces in $\mathbb{R}^{n}$. We explicitly construct the generalized bitwise operators involved in this process by making use of the framework of digital calculus, that we have very recently introduced \cite{VGM4, QUANTUM, CHAOSOLFRAC}. We show that these abstract surfaces give rise to a roughness exponent $H$ that shapes the resulting spatial patterns, larger values of the exponent leading to coarser surfaces. 

Recently, we have found a new construction for fractals based on a fractal decomposition of a given function \cite{CHAOSOLFRAC} linking the resulting fractal objects to finite group theory \cite{CHAOSOLFRAC}. The construction of fractal surfaces given here is new and different to our previous one in \cite{CHAOSOLFRAC}. Although operators acting on two variables are again considered, these operators are not restricted to have the latin square property as in \cite{CHAOSOLFRAC} and the construction is different even when the concepts that are used in it are closely related.

In Section \ref{B} we introduce the main concept of digital calculus, the digit function \cite{QUANTUM, CHAOSOLFRAC} and we construct all discrete operators acting on two variables whose values are restricted to be any of the $p$ integers in the interval $[0,p-1]$. With the aid of these tools, we then construct local bitwise operators acting on points in $\mathbb{R}^{N}$ and we rigorously prove that all give rise to (hyper)surfaces satisfying the self-affine property meant by Eq. (\ref{selfafi}). We then also prove that the roughness exponent behaves as in experimental physical systems, larger values of it leading to coarser surfaces \cite{Sinha,Chiarello,Indekeu}. In Section \ref{Bit} we illustrate these results with specific numerical examples. 

\section{Digital calculus: The digit function, discrete operators and generalized bitwise arithmetic} \label{B}

In this article $S$ denotes the set of $p$ integers in the interval $[0,p-1]$. The digit function $\mathbf{d}_{p}(k,x)$, for $p \in \mathbb{N}$, $k \in \mathbb{Z}$ and $x \in \mathbb{R}$, is the surjective mapping $\mathbb{R} \to S$ defined as \cite{QUANTUM, CHAOSOLFRAC, VGM4} 
\begin{equation}
\mathbf{d}_{p}(k,x)=\left \lfloor \frac{x}{p^{k}} \right \rfloor-p\left \lfloor \frac{x}{p^{k+1}} \right \rfloor    \label{cucuAreal}
\end{equation}
and gives the $k$-th digit of the real number $x$ (when it is non-negative) in a positional numeral system in radix $p > 1$. If $p=1$ the digit function satisfies $\mathbf{d}_{1}(k,x)=\mathbf{d}_{1}(0,x)=0$ and it does not relate to a positional numeral system. In Eq. (\ref{cucuAreal}) $\lfloor \ldots \rfloor$ denotes the floor function (lower closest integer) of the quantity between the brackets.

With the digit function, we can express any real number $x$ as \cite{QUANTUM, CHAOSOLFRAC, VGM4}  
\begin{equation}
x=\text{sign}(x)\sum_{k=-\infty}^{\lfloor \log_{p}|x| \rfloor} p^{k} \mathbf{d}_{p}(k,|x|) \label{idenreal}
\end{equation}
The integer part of $x$ is given by 
\begin{equation}
\text{sign}(x) \lfloor |x| \rfloor =\text{sign}(x)\sum_{k=0}^{\lfloor \log_{p}|x| \rfloor} p^{k} \mathbf{d}_{p}(k,|x|) \label{idenreal2}
\end{equation}

\noindent \emph{Example:} In the decimal radix $p=10$, the number $\pi \equiv 3.1415\ldots$ has digits $\mathbf{d}_{10}(0,\pi)=3$, $\mathbf{d}_{10}(-1,\pi)=1$, $\mathbf{d}_{10}(-2,\pi)=4$, $\mathbf{d}_{10}(-3,\pi)=1$, $\mathbf{d}_{10}(-4,\pi)=5$, etc.

In general, a truncation to $D$ digits after the radix point is given by
\begin{equation}
\text{sign}(x) p^{-D}\lfloor p^{D}|x| \rfloor =\text{sign}(x)\sum_{k=-D}^{\lfloor \log_{p}|x| \rfloor} p^{k} \mathbf{d}_{p}(k,|x|) \label{idenrealcog}
\end{equation}
We shall call $p^{-D}\lfloor p^{D}\ldots \rfloor$, with $D$ being an integer number, the \emph{coarse graining operator}.

\begin{figure*} 
\includegraphics[width=0.8 \textwidth]{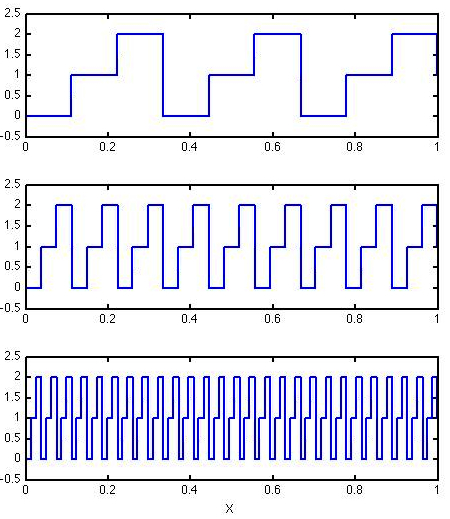}
\caption{\scriptsize{The digit function $\mathbf{d}_{3}(k,x)$ plotted for $k=-2, -3, -4$ (from top to bottom) in the interval $x \in [0,1]$. \cite{arxiv2}}} \label{digitfu}
\end{figure*} 

Most properties of the digit function are easily understood from its plot, as shown in Fig. \ref{digitfu} for the case $p=3$ and $k=-2,-3,-4$. The digit function $\mathbf{d}_{p}(k,x)$ is in all cases a periodic staircase that is set to zero each time that $x$ is an integer multiple of $p^{k+1}$. The digit function possesses an important scaling property \cite{CHAOSOLFRAC} that we shall use below. We have
\begin{eqnarray}
\mathbf{d}_{p}(k,p^{m}x)&=&\left \lfloor \frac{p^{m}x}{p^{k}} \right \rfloor-p\left \lfloor \frac{p^{m}x}{p^{k+1}} \right \rfloor = \left \lfloor \frac{x}{p^{k-m}} \right \rfloor-p\left \lfloor \frac{x}{p^{k-m+1}} \right \rfloor= \mathbf{d}_{p}(k-m,x) \nonumber \\ 
&& \label{scal}
\end{eqnarray} 

Let $n$ be a natural number. Then, the following relationship also holds
\begin{eqnarray}
\mathbf{d}_{np}(k,x)&=&\mathbf{d}_{p}\left(k, \frac{x}{n^{k}} \right)+p\mathbf{d}_{n}\left(k, \frac{x}{p^{k+1}} \right)=\mathbf{d}_{n}\left(k, \frac{x}{p^{k}} \right)+n\mathbf{d}_{p}\left(k, \frac{x}{n^{k+1}} \right)    \label{reldi1} 
\end{eqnarray}
This can be easily seen by using the definition, since we have
\begin{eqnarray}
\mathbf{d}_{p}\left(k, \frac{x}{n^{k}} \right)+p\mathbf{d}_{n}\left(k, \frac{x}{p^{k+1}} \right)&=&
\left \lfloor \frac{x}{p^{k}n^{k}} \right \rfloor -p \left \lfloor \frac{x}{p^{k+1}n^{k}} \right \rfloor
+p \left \lfloor \frac{x}{p^{k+1}n^{k}} \right \rfloor-np\left \lfloor \frac{x}{p^{k+1}n^{k+1}} \right \rfloor \nonumber \\
&=& \left \lfloor \frac{x}{(np)^{k}} \right \rfloor -np\left \lfloor \frac{x}{(np)^{k+1}} \right \rfloor =\mathbf{d}_{np}(k,x)
\end{eqnarray}
which proves Eq. (\ref{reldi1}).

Let us now construct the operators $+_{p}$ and $\hat{+}_{p}$ mentioned in the introduction. We consider $a$ and $b$ nonnegative real numbers, for simplicity. From the digit function, Eq. (\ref{idenreal}), we have that, on one hand
\begin{equation}
a+b=\sum_{k=-\infty}^{\lfloor \log_{p}(a+b) \rfloor} p^{k} \mathbf{d}_{p}(k,a+b) \label{idensum1}
\end{equation}
and, on the other hand,
\begin{equation}
a+b=\sum_{k=-\infty}^{\lfloor \log_{p}a \rfloor} p^{k}\mathbf{d}_{p}(k,a) +\sum_{k=-\infty}^{\lfloor \log_{p}b \rfloor} p^{k} \mathbf{d}_{p}(k,b)=\sum_{k=-\infty}^{\max \{ \lfloor \log_{p}a \rfloor,\lfloor \log_{p}b \rfloor \} } p^{k}\left(\mathbf{d}_{p}(k,a) + \mathbf{d}_{p}(k,b)\right) \label{idensum2}
\end{equation}
Since $p^{2} > \mathbf{d}_{p}(k,a) + \mathbf{d}_{p}(k,b)$, we have $\mathbf{d}_{p^{2}}\left(0,\mathbf{d}_{p}(k,a) + \mathbf{d}_{p}(k,b)\right)=\mathbf{d}_{p}(k,a) + \mathbf{d}_{p}(k,b)$  and we can write this last expression as
\begin{equation}
a+b=\sum_{k=-\infty}^{\max \{ \lfloor \log_{p}a \rfloor,\lfloor \log_{p}b \rfloor \} } p^{k}\mathbf{d}_{p^{2}}\left(0,\mathbf{d}_{p}(k,a) + \mathbf{d}_{p}(k,b)\right) \label{idensum3}
\end{equation}
Therefore, by using Eq. (\ref{reldi1})
\begin{equation}
a+b=\sum_{k=-\infty}^{\max \{ \lfloor \log_{p}a \rfloor,\lfloor \log_{p}b \rfloor \} } p^{k}\mathbf{d}_{p}\left(0,\mathbf{d}_{p}(k,a) + \mathbf{d}_{p}(k,b)\right)+\sum_{k=-\infty}^{\max \{ \lfloor \log_{p}a \rfloor,\lfloor \log_{p}b \rfloor \} } p^{k+1}\mathbf{d}_{p}\left(1,\mathbf{d}_{p}(k,a) + \mathbf{d}_{p}(k,b)\right) \label{idensum4}
\end{equation}
Thus, we see that the ordinary sum of two real numbers can be split into the bitwise addition modulo $p$ of $a$ and $b$
\begin{equation}
a+_{p}b=\sum_{k=-\infty}^{\max \{ \lfloor \log_{p}a \rfloor,\lfloor \log_{p}b \rfloor \} } p^{k}\mathbf{d}_{p}\left(0,\mathbf{d}_{p}(k,a) + \mathbf{d}_{p}(k,b)\right) \label{mpo}
\end{equation}
plus a term which provides the total contribution of the carries
\begin{equation}
a\hat{+}_{p}b=\sum_{k=-\infty}^{\max \{ \lfloor \log_{p}a \rfloor,\lfloor \log_{p}b \rfloor \} } p^{k+1}\mathbf{d}_{p}\left(1,\mathbf{d}_{p}(k,a) + \mathbf{d}_{p}(k,b)\right)
\end{equation}
Note that if the kth-digits of $a$ and $b$ are such that $\mathbf{d}_{p}(k,a) + \mathbf{d}_{p}(k,b)<p$ there is no contribution to the carries coming from the $k$-th digit since in that case $\mathbf{d}_{p}\left(1,\mathbf{d}_{p}(k,a) + \mathbf{d}_{p}(k,b)\right)=0$. We thus have proved
\begin{equation}
a+b=(a+_{p}b)+(a\hat{+}_{p}b)
\end{equation}
Our goal in this paper is to find all possible bitwise operators which, as is the case of $+_{p}$, act on two arbitrary real numbers.  This goal is attained by first constructing all discrete operators that are laws of composition (magmas) of $S\times S \to S$. We denote such operators by $_{2}R_{p}(x_{0},x_{1})$. They act on integers $x_{0}$ and $x_{1}$, both in $S$. The operator can be defined through a table as
\begin{center}
\begin{tabular}{c|ccccc}
%\hline 
$ _{2}R_{p} $ & $0$ & $1$ & $\ldots k \ldots$ & $p-2$ & $p-1$ \\
\hline
$0 $ & $a_{0}$ & $a_{1}$  & $\ldots a_{k} \ldots $ & $a_{p-2}$ & $a_{p-1}$ \\
$1 $ & $a_{p}$ & $a_{p+1}$  & $\ldots a_{p+k} \ldots $ & $a_{2p-2}$ & $a_{2p-1}$ \\
$2 $ & $a_{2p}$ & $a_{2p+1}$ & $\ldots a_{2p+k} \ldots $ & $a_{3p-2}$ & $a_{3p-1}$ \\
$\ldots k \ldots $ & $a_{kp}$ & $a_{kp+1}$  & $\ldots a_{kp+k} \ldots $ & $a_{(k+1)p-2}$ & $a_{(k+1)p-1}$ \\
$\ p-2\ $ & $\ a_{p(p-2)}\ $ & $a_{p(p-2)+1}\ $  & $\ldots a_{p(p-2)+k} \ldots $ & $a_{p(p-1)-2}$ & $\ a_{p(p-1)-1}\ $ \\
$p-1 $ & $\ a_{p(p-1)}\ $ & $a_{p(p-1)+1}\ $  & $\ldots a_{p(p-1)+k} \ldots $ & $\ a_{p^{2}-2}\ $ & $a_{p^{2}-1}$ \\
%\hline
\end{tabular}
\end{center}
where $x_{0}$ is listed in the columns and $x_{1}$ in the rows, and the result $a_{x_{0}+px_{1}} \in S$,  is given by the table. Clearly, we have for any $a_{n\in \mathbb{N}}$ that $n \in [0, p^{2}-1]$. Since all $a_{n}$'s are integers $\in [0,p-1]$ this table provides the most general description of a law of composition (also called a magma) \cite{Lang, Bruck}. The integer number $R$ entering in the label of the operator $_{2}R_{p}$ is defined as
\begin{equation}
R \equiv \sum_{n=0}^{p^{2}-1}a_{n}p^{n}. \label{RWolf}
\end{equation}
Thus, $R \in [0,p^{p^{2}}-1]$. Therefore, it is clear that, from the above table
\begin{equation}
_{2}R_{p}(x_{0},x_{1})=a_{x_{0}+px_{1}}=\mathbf{d}_{p}\left(x_{0}+px_{1}, \sum_{n=0}^{p^{2}-1}p^{n}a_{n} \right)=\mathbf{d}_{p}\left(x_{0}+px_{1},R \right) \label{concise2}
\end{equation}
since the $(x_{0}+px_{1})$-th digit of the number $\sum_{n=0}^{p^{2}-1}p^{n}a_{n}$ is $a_{x_{0}+px_{1}}$. This last expression in terms of the digit function most concisely expresses any possible magma. Let now $u$ and $v$ be arbitrary real numbers. The generalized bitwise operator $\mathbf{b}_{q}\left(_{2}R_{p}; u, v \right)$, with $q\ge 2$ acting on these real numbers is defined as  
\begin{eqnarray}
\mathbf{b}_{q}\left(_{2}R_{p}; u, v\right) &\equiv& \sum_{k=-\infty}^{k_{max} }q^{k}\ _{2}R_{p}\left(\mathbf{d}_{p}(k,u), \mathbf{d}_{p}(k,v)\right) \nonumber \\
&=& \sum_{k=-\infty}^{k_{max}} q^{k} \mathbf{d}_{p}\left(\mathbf{d}_{p}(k,u)+p\mathbf{d}_{p}(k,v), R \right) \label{bwq}
\end{eqnarray}
where $k_{max}=\max \{\lfloor \log_{p}u \rfloor, \lfloor \log_{p}v \rfloor \}$. This is the expression that we shall analyze in the rest of the article. 

\noindent \emph{Example:} Let us find what is the numerical value of $\mathbf{b}_{2}\left(_{2}14_{2}; 5, 11\right)$. We have that 5 and 11 are given respectively in base $p=2$ as '0101' and '1011'. Now the operator $_{2}14_{2}$ has vector $(a_{0},a_{1},a_{2},a_{3})=(0,1,1,1)$ which means that it returns one when the digit $k$ of $u$ or $v$ or both is one and zero otherwise. Then, the resulting string is $(1\ 1\ 1\ 1)$ and, therefore, $\mathbf{b}_{q}\left(_{2}14_{2}; 5, 11\right)=1\cdot q^{0}+1\cdot q^{1}+1\cdot q^{2}+1\cdot q^{3}$. Thus $\mathbf{b}_{2}\left(_{2}14_{2}; 5, 11\right)=15$.

Let us now also see how the operator $+_{p}$ is a particular instance of Eq. (\ref{bwq}). We note that if we take $a_{n}=\mathbf{d}_{p}\left(0, \mathbf{d}_{p}\left(0, n \right)+\mathbf{d}_{p}\left(1, n \right) \right)$ so that
\begin{equation}
R=\sum_{n=0}^{p^{2}-1}p^{n}\mathbf{d}_{p}\left(0, \mathbf{d}_{p}\left(0, n \right)+\mathbf{d}_{p}\left(1, n \right) \right) \label{Rejem}
\end{equation}
we then have, from Eq. (\ref{concise2}) 
\begin{eqnarray}
&&\mathbf{d}_{p}\left(\mathbf{d}_{p}(k,u)+p\mathbf{d}_{p}(k,v), R \right)=\mathbf{d}_{p}\left(\mathbf{d}_{p}(k,u)+p\mathbf{d}_{p}(k,v), \sum_{n=0}^{p^{2}-1}p^{n}\mathbf{d}_{p}\left(0, \mathbf{d}_{p}\left(0, n \right)+\mathbf{d}_{p}\left(1, n \right) \right) \right)\nonumber \\
&&=\mathbf{d}_{p}\left(0, \mathbf{d}_{p}\left(0, \mathbf{d}_{p}(k,u)+p\mathbf{d}_{p}(k,v) \right)+\mathbf{d}_{p}\left(1, \mathbf{d}_{p}(k,u)+p\mathbf{d}_{p}(k,v)\right) \right)=\mathbf{d}_{p}\left(0, \mathbf{d}_{p}(k,u)+p\mathbf{d}_{p}(k,v) \right) \nonumber
\end{eqnarray}
Therefore, by taking $q=p$ in Eq. (\ref{bwq}) and replacing this expression we obtain
\begin{eqnarray}
\mathbf{b}_{p}\left(_{2}R_{p}; u, v\right) &=& \sum_{k=-\infty}^{k_{max}} p^{k} \mathbf{d}_{p}\left(\mathbf{d}_{p}(k,u)+p\mathbf{d}_{p}(k,v), R \right)=\sum_{k=-\infty}^{k_{max}} p^{k}\mathbf{d}_{p}\left(0, \mathbf{d}_{p}(k,u)+p\mathbf{d}_{p}(k,v) \right) \nonumber \\
&=& u+_{p}v
\end{eqnarray}
as can be seen from Eq. (\ref{mpo}). Thus, the addition modulo $p$ of two real numbers $u$,  $v$ is a particular instance of a generalized bitwise arithmetic operator $\mathbf{b}_{q}\left(_{2}R_{p}; u, v\right)$ with $q=p$ and with the magma $_{2}R_{p}$ having code $R$ given by Eq. (\ref{Rejem}).

We note that our expression Eq. (\ref{concise2}) for any magma in terms of the digit function can be generalized to an arbitrary $N$-ary operator $_{N}R_{p}: S^{N}\to S$ as

 \begin{equation}
_{N}R_{p}(x_{0},\ldots, x_{N-1})\equiv \mathbf{d}_{p}\left(\sum_{k=0}^{N-1}p^{k}x_{k}, \sum_{n=0}^{p^{N}-1}p^{n}a_{n} \right)=\mathbf{d}_{p}\left(\sum_{k=0}^{N-1}p^{k}x_{k}, R \right) \label{concise}
\end{equation}
where now $R=\sum_{n=0}^{p^{N}-1}a_{n}p^{n}$ is an integer $\in [0, p^{p^{N}}-1]$. This operator can be used to define a generalized bitwise operator acting on $N$ real numbers $u_{0}$, $u_{1}$, $\ldots$, $u_{N-1}$ as
\begin{eqnarray}
\mathbf{b}_{q}\left(\ _{N}R_{p}; u_{0},\ldots, u_{N-1}\right) &\equiv& \sum_{k=-\infty}^{\max \{\lfloor \log_{p}{u_{0}}  \rfloor, \ldots, \lfloor \log_{p}{u_{N-1}} \rfloor \} } \ _{N}R_{p}\left(\mathbf{d}_{p}(k,u_{0}), \ldots , \mathbf{d}_{p}(k,u_{N-1})\right)q^{k} \nonumber \\ 
&=&\sum_{k=-\infty}^{\max \{\lfloor \log_{p}{u_{0}}  \rfloor, \ldots, \lfloor \log_{p}{u_{N-1}} \rfloor \}} q^{k}\mathbf{d}_{p}\left(\sum_{n=0}^{N-1}p^{n}\mathbf{d}_{p}(k,u_{n}), R \right)
\label{bwqN}
\end{eqnarray}
This function acts in the following way: it takes the digit $k$ of each number $u_{0}$,$\ldots$, $u_{N-1}$ in radix $p$ and introduces the N-ary operator $_{N}R_{p}$ given by Eq. (\ref{concise}) on the $N$ digits. Then, the resulting string of digits is interpreted as a number in radix $q$ (if $\ge p$) or simply, in general, as a polynomial in powers of $q$. Thus, $\mathbf{b}_{q}$ is a non-negative real number.

We now prove one of the main results of this article.

\textbf{Theorem 2.1 (Self-affinity of the generalized bitwise function.)} \emph{The function $\mathbf{b}_{q}$ defined by Eq. (\ref{bwqN}) has a self-affine property, so that}
\begin{equation}
\mathbf{b}_{q}\left(\ _{N}R_{p}; u_{0},\ldots, u_{N-1}\right)= p^{-\ln q}\mathbf{b}_{q}\left(\ _{N}R_{p}; pu_{0},\ldots, pu_{N-1}\right) \label{afibwq}
\end{equation}	
\emph{so that $H=\log_{p}q$ defines a 'roughness exponent' of a hypersurface in $N$ dimensions, cf. Eq. (\ref{selfafi}).}

\noindent \emph{Proof:} We have
\begin{eqnarray}
&&\mathbf{b}_{q}\left(\ _{N}R_{p}; pu_{0},\ldots, pu_{N-1}\right) = \sum_{k=-\infty}^{\max \{\lfloor \log_{p}{pu_{0}}  \rfloor, \ldots, \lfloor \log_{p}{pu_{N-1}} \rfloor \}}\ _{N}R_{p}\left(\mathbf{d}_{p}(k,pu_{0}),\ldots, \mathbf{d}_{p}(k,pu_{N-1})\right)q^{k}  \nonumber \\
&&= \sum_{k=-\infty}^{1+\max \{\lfloor \log_{p}{u_{0}}  \rfloor, \ldots, \lfloor \log_{p}{u_{N-1}} \rfloor \}}\ _{N}R_{p}\left(\mathbf{d}_{p}(k-1,u_{0}),\ldots, \mathbf{d}_{p}(k-1,u_{N-1})\right)q^{k} \nonumber \\
&&= \sum_{k'=-\infty}^{\max \{\lfloor \log_{p}{u_{0}}  \rfloor, \ldots, \lfloor \log_{p}{u_{N-1}} \rfloor \}}\ _{N}R_{p}\left(\mathbf{d}_{p}(k',u_{0}),\ldots, \mathbf{d}_{p}(k',u_{N-1})\right)q^{k'+1}  \nonumber \\
&&= p^{\log_{p}q} \sum_{k'=-\infty}^{\max \{\lfloor \log_{p}{u_{0}}  \rfloor, \ldots, \lfloor \log_{p}{u_{N-1}} \rfloor \}}\ _{N}R_{p}\left(\mathbf{d}_{p}(k',u_{0}),\ldots, \mathbf{d}_{p}(k',u_{N-1})\right)q^{k'} \nonumber \\
&=& p^{\log_{p}q} \mathbf{b}_{q}\left(\ _{N}R_{p}; u_{0},\ldots, u_{N-1}\right) 
\end{eqnarray}
where we have used the scaling property, Eq. (\ref{scal}) in noting that
\begin{equation}
\mathbf{d}_{p}(k,px)=\mathbf{d}_{p}(k-1,x)
\end{equation}
This proves the result and that  $\log_{p}q=H$ is the roughness exponent of the self-affine surface by comparing this latter expression with Eq. (\ref{selfafi}). $\Box$

The crucial property established in general by the above theorem proves that the operator $x+_{p}y$ in Fig. \ref{dsums} has indeed a self-affine property and deserves to be called a fractal in its own right.

We also have the following result.

\textbf{Theorem 2.2 (Coarse-graining theorem).} \emph{Let $_{N}R_{p}$ be a N-variable discrete operator in radix $p$ acting bitwise on the digits of the reals $u_{0}$, $u_{1}$, $\ldots$, $u_{N-1}$. Then, the function $\mathbf{b}_{q}\left(R_{p}; u_{0},\ u_{1},\ \ldots,\ u_{N-1}\right)$ defined in Eq. (\ref{bwqN}) is smoothened as $q$ is increased. Furthermore, if $q >> p$ the following expression is asymptotically satisfied}
\begin{equation}
\mathbf{b}_{q} \sim \left \lfloor \frac{\mathbf{b}_{p}}{p^{k_{max}}} \right \rfloor q^{k_{max}} \label{asymtot}
\end{equation}
\emph{where $k_{max}=\max \{\lfloor \log_{p}{u_{0}} \rfloor, \lfloor \log_{p}{u_{1}} \rfloor, \ldots, \lfloor \log_{p}{u_{N-1}} \rfloor \}$.}

\noindent \emph{Proof:} Since $\mathbf{b}_{q}$ is non-negative by definition, from Eq. (\ref{idenreal}) we have $\mathbf{b}_{q}=\sum_{k=-\infty}^{k_{max}} q^{k} \mathbf{d}_{q}(k,\mathbf{b}_{q})$. All coefficients accompanying powers of $q$ in Eq. (\ref{bwq}) are non-negative integers. Thus $\mathbf{b}_{q}$ increases monotonically with $q$. Since all coefficients are bounded (and all are independent of $q$) and their dependence on the coordinates $u_{n}$, $n=0,\ldots, N-1$ is governed by $k$ it is clear then that, as $q$ increases, terms with $k$ larger become more significant and the sum displays a longer spatial variability. As a consequence of this, details are gradually lost and coarser surfaces are obtained. Now, note also that $ \mathbf{b}_{p}=\sum_{k=-\infty}^{k_{max}} p^{k} \mathbf{d}_p(k,\mathbf{b}_{p})$. When $q>>p$ is large only the most significant digit of $\mathbf{b}_{q}$ becomes relevant and we have
\begin{equation}
\mathbf{b}_{q} \sim q^{k_{max}} \mathbf{d}_{q}(k_{max},\mathbf{b}_{q}) \label{asympro}
\end{equation}
Now, since from Eq. (\ref{bwqN}), for $q \ge p$ we have
\begin{equation}
\mathbf{d}_{q}(k,\mathbf{b}_{q})=\mathbf{d}_p(k,\mathbf{b}_{p})=\mathbf{d}_{p}\left(\sum_{n=0}^{N-1}p^{n}\mathbf{d}_{p}(k,u_{n}), R \right) 
\end{equation}
and the digits of $\mathbf{b}_{q}$ are \emph{the same} as those of $\mathbf{b}_{p}$ and, hence, belong to the set $S$. Then, from Eq. (\ref{asympro}) and, by using the definition of the digit function, Eq. (\ref{cucuAreal}) and noting that $\lfloor \mathbf{b}_{p}/p^{k_{max}+1} \rfloor =0$, we finally obtain
\begin{equation}
\mathbf{b}_{q} \sim q^{k_{max}} \mathbf{d}_{q}(k_{max},\mathbf{b}_{q})=
q^{k_{max}} \mathbf{d}_{p}(k_{max},\mathbf{b}_{p})=
 \left \lfloor \frac{\mathbf{b}_{p}}{p^{k_{max}}} \right \rfloor q^{k_{max}}  
 \label{asympro2}
\end{equation}
which proves the result. $\Box$

The coarse graining theorem implies that $\mathbf{b}_{q}$ \emph{asymptotically} behaves like the most significant digit of $\mathbf{b}_{p}$ for $q>>p$, after a trivial rescaling (i.e. dividing by $q^{k_{max}}$). Since all information contained in the less significant digits of $\mathbf{b}_{p}$ is lost in $\mathbf{b}_{q}$, the latter exhibits no fine details compared to the former and if one takes the numbers $u_{0}$, $u_{1}$, $\ldots$, $u_{N-1}$ as 'coordinates'  only the broadest patches in the space spanned along these coordinates survive: Only the operations on the most significant digit of the numbers $u_{0}$, $u_{1}$, $\ldots$, $u_{N-1}$ become relevant.

\section{Patchwork quilts: Self-affine surfaces and the roughness exponent} \label{Bit}

\begin{figure}
\includegraphics[width=1.0 \textwidth]{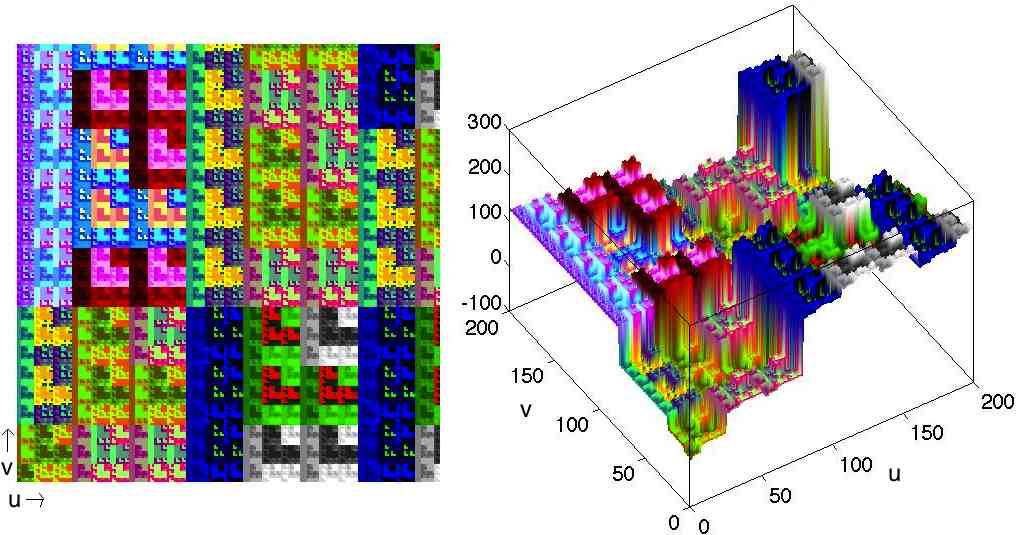}
\begin{center}
\caption{\scriptsize{Two views of the function $\mathbf{b}_{3}\left(_{2}13903_{3}; u, v \right)$ in a portion of the plane with $u \in [0,200]$ and $v \in [0,200]$}}.  \label{patch3D}
\end{center}
\end{figure}
\begin{figure}
\includegraphics[width=0.8 \textwidth]{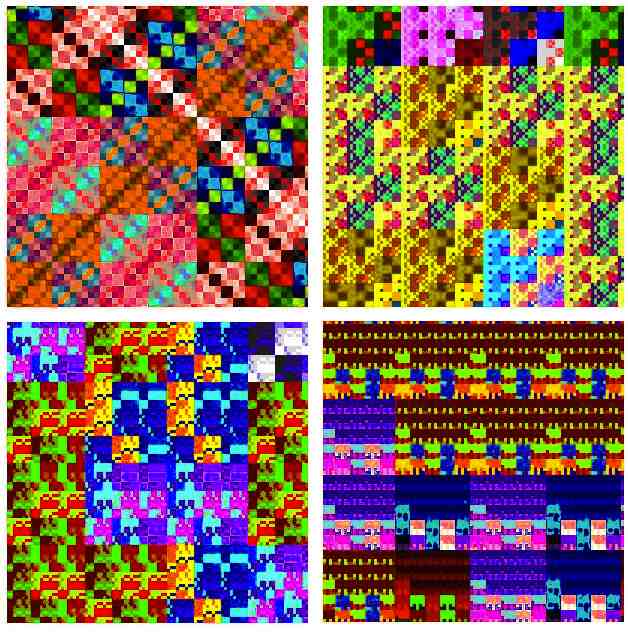}
\begin{center}
\caption{\scriptsize{From top to bottom and left to right, the functions $\mathbf{b}_{2}\left(_{2}6_{2}; u, v \right)$, $\mathbf{b}_{3}\left(_{2}7417_{3}; u, v \right)$, $\mathbf{b}_{3}\left(_{2}9407_{3}; u, v \right)$ and $\mathbf{b}_{5}\left(_{2}13427417_{5}; u, v \right)$, with $u \in [0,100]$ and $v \in [0,100]$}}. \label{patch1}
\end{center}
\end{figure}

To better substantiate these results and study particular examples, it is better to fix $N=2$ in Eq. (\ref{bwqN}) i.e., to work with Eq. (\ref{bwq}), so that we can represent such operators as functions taking values on the plane given by coordinates $u$ and $v$.

In Figs. \ref{patch3D} and \ref{patch1} $\mathbf{b}_{q}$ is plotted as a function of $(u,v)$ and the operator indicated in each case. Patchwork quilts are obtained in every case when projecting the 3D object on the plane (the figures just only show a finite portion of the plane). 
Such representations on the plane are useful to get a glimpse of the behavior of the operator $_{2}R_{p}$ within $\mathbf{b}_{q}$. For example that $_{2}6_{2}$ is commutative can be clearly seen from the reflection axis on the southwest-northeast diagonal exhibited by $\mathbf{b}_{2}\left(_{2}6_{2}; u, v \right)$ (see Fig. \ref{patch1}, top left). The different patches give information on how $\mathbf{b}_{q}$ organizes in the plane. The tridimensional aspect of these surfaces, an example is shown in Fig. \ref{patch1}, reminds the towers of the 'exotic' fractal surfaces reported by Indekeu et al. obtained through random deposition \cite{Indekeu5}.

\begin{center}
\begin{figure}
\includegraphics[width=0.8 \textwidth]{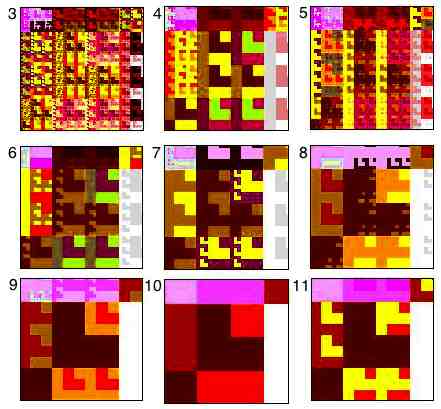}
\begin{center}
\caption{\scriptsize{Gradual spatial coarsening of the bitwise operation function $\mathbf{b}_{q}$ as the parameter $q$ is increased. Shown is the function $\mathbf{b}_{q}\left(_{2}13903_{3}; u, v \right)$ for all integer values of $q \in [3,11]$ as indicated besides each panel ($u \in [0,100]$ and $v \in [0,100]$). }} \label{coarse}
\end{center}
\end{figure}
\end{center}

The roughness exponent obtained above has some of the nice properties reported in experimental studies on fractal self-affine surfaces \cite{Sinha} \cite{Chiarello} \cite{Indekeu}: A larger value of the exponent corresponds to a coarser (i.e. 'smoother') surface. This is a result of the coarse graining theorem in the previous section, which we illustrate in Fig. \ref{coarse} for the case $q>p$ (the case $q\le p$ follows a similar trend although not that markedly).
In Fig. \ref{coarse}, the function $\mathbf{b}_{q}\left(13903_{3}; u, v \right)$ is plotted for the values of $q \in [3,11]$ $(p=3)$ shown in the Figure. We observe that for $q$ increasingly large, details are gradually lost and only the coarser structures survive. In the Figure it is observed that for $q=10$ the asymptotic limit is already being approached. Remarkably, for $q$ prime details sometimes reemerge (but in a coarser scale).  This rather subtle phenomenon poses an interesting number theoretical problem for which we have not yet a solution, and which shall be addressed elsewhere.

Increasing $q$ leads to coarser structures. If $q$ is not asymptotically large the structures at  different scales overlap creating complex patterns that are coarser on the average. We can then ask the question whether is it possible to sharply discriminate the coarser structures without significantly changing the value of the function (beyond, of course, the change induced by neglecting the fine details). This is accomplished by the coarse graining operator acting on $\mathbf{b}_{q}$, i.e. $q^{-D}\lfloor q^{D}\mathbf{b}_{q}\left(_{2}R_{p}; u, v\right) \rfloor$. By applying this operator to  Eq. (\ref{bwq}) we obtain
\begin{eqnarray}
q^{-D}\lfloor q^{D}\mathbf{b}_{q}\left(_{2}R_{p}; u, v\right) \rfloor
&=& \sum_{k=-D}^{k_{max}} q^{k} \mathbf{d}_{p}\left(\mathbf{d}_{p}(k,u)+p\mathbf{d}_{p}(k,v), R \right) \label{bwqtrunc}
\end{eqnarray}
The error committed in approximating $\mathbf{b}_{q}$ by such truncation is thus strictly smaller than $q^{-D}$. Since each term accompanying a power $k$ of $q$ contributes a spatial structure (that is coarser for $k$ larger), making $D$ more negative selects the coarser structures while making the value of $q^{-D}\lfloor q^{D}\mathbf{b}_{q}\left(_{2}R_{p}; u, v\right) \rfloor$ to slightly \emph{decrease}.

\begin{center}
\begin{figure}
\includegraphics[width=1.0 \textwidth]{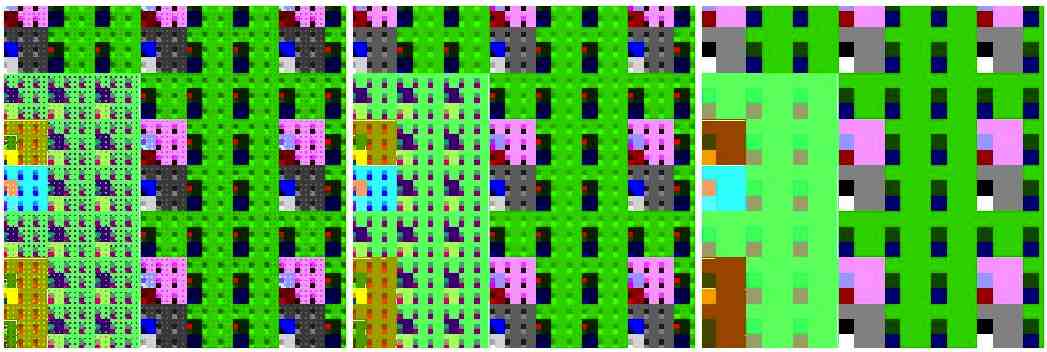}
\begin{center}
\caption{\scriptsize{Action of the coarse graining operator on a self-affine surface, as described by Eq. (\ref{bwqtrunc}). Shown is the quantity $3^{-D}\lfloor 3^{D}\mathbf{b}_{3}\left(_{2}9815_{3}; u, v\right) \rfloor$ for the values $D=0$ (left), $D=-1$ (center) and $D=-2$ (right) with $u \in [0,100]$ and $v \in [0,100]$.}} \label{coarsope}
\end{center}
\end{figure}
\end{center}

The action of the coarse graining operator, as described by Eq. (\ref{bwqtrunc}), is shown in Fig. \ref{coarsope} for the function $3^{-D}\lfloor 3^{D}\mathbf{b}_{3}\left(_{2}9815_{3}; u, v\right) \rfloor$ for the values $D=0$ (left), $D=-1$ (center) and $D=-2$ (right). We see that the coarse graining operator merely selects broader patches, eliminating the fine details. 

The structure of the coarse graining operator is also interesting in the sense that it reveals that the operation of multipliying by an scalar, i.e. $q^{-D}$, and taking the floor operator $\left \lfloor \ldots \right \rfloor$ do not generally commute: $q^{-D}\lfloor \ldots \rfloor \ne \lfloor q^{-D} \ldots \rfloor$. This lack of commutativity is behind the loss of information implied by the coarse graining. In \cite{QUANTUM} we have discussed the implications that, in our view, this fact has for the foundations of quantum mechanics.

\section{Conclusions}

In this article a new general method to mathematically design fractal surfaces with self-affine properties has been presented. We have discovered these structures hidden within operations as simple as the ordinary sum of real numbers. Thus, we speculate that such structures may be of general interest in the mathematical modeling of complex physical systems. Our approach is based on digital calculus, which we have introduced to describe rule-based dynamical systems, as cellular automata \cite{VGM1} and substitution systems \cite{VGM4}. This approach is also the cornerstone of a recent formulation of quantum mechanics \cite{QUANTUM} and of our method to derive fractal decompositions of conserved physical quantities \cite{CHAOSOLFRAC}. We have rigorously proved that bitwise operators acting on real scalar fields on each point in the plane $\mathbb{R}^{2}$ lead to surfaces with self-affine properties. We have obtained the roughness exponent for these surfaces and shown that it has the characteristic nice properties expected for such an exponent, as widely reported in experimental systems: The larger this exponent, the coarser the resulting surface  \cite{Sinha, Chiarello, Indekeu}.

The fractal surfaces here obtained come under the general name of 'patchwork quilts' because we are following the acknowledgment made by Donald Knuth in \cite{Knuth} to one pattern designed by D. Sleator in 1976 (unpublished result, displayed on p. 136 of \cite{Knuth}) resembling the ones obtained here. However, no relationship of such pattern to fractals and self-affinity seems to have ever been attempted and no systematic construction of such patterns is found in the literature. In this article, Eq. (\ref{bwqN}) constitutes a universal form embodying all conceivable bitwise operators in any radices $p$ and $q$ of standard numeral systems and any number of real scalars $N$, the different digits representing independent bits of information at different scales. 

We speculate that our surfaces can be related to free energy landscapes of clusters and biomolecules \cite{Wales} obtained under the basin hopping scheme \cite{WalesDoye}. An analysis of this connection shall be investigated elsewhere.

\bibliography{biblos}{}
\bibliographystyle{h-physrev3.bst}
%\bibliographystyle{spbasic}
%\bibliographystyle{spphys}

%\begin{thebibliography}{}
%%
%% and use \bibitem to create references. Consult the Instructions
%% for authors for reference list style.
%\end{thebibliography}

\end{document}